\documentclass[aps,twocolumn,superscriptaddress,a4paper]{revtex4}
\usepackage{graphicx,amsmath,verbatim, hyperref}
\usepackage{upgreek}
\usepackage{capt-of}
\usepackage{lmodern}

\begin{document}
\title{A Scanning Cavity Microscope}
\author{Matthias Mader}
\affiliation{Ludwig-Maximilians-Universit\"at M\"unchen, Fakult\"at f\"ur Physik, Schellingstra\ss e 4, 80799 M\"unchen, Germany}
\affiliation{Max-Planck-Institut f\"ur Quantenoptik, Hans-Kopfermann-Stra\ss e 1, 85748 Garching, Germany}

\author{Jakob Reichel}
\affiliation{Laboratoire Kastler Brossel, ENS/UPMC-Paris 6/CNRS, 24 rue Lhomond, F-75005 Paris, France}

\author{Theodor W. H\"ansch}
\affiliation{Ludwig-Maximilians-Universit\"at M\"unchen, Fakult\"at f\"ur Physik, Schellingstra\ss e 4, 80799 M\"unchen, Germany}
\affiliation{Max-Planck-Institut f\"ur Quantenoptik, Hans-Kopfermann-Stra\ss e 1, 85748 Garching, Germany}

\author{David Hunger}
\email[To whom correspondence should be addressed. E-mail: ]{david.hunger@physik.lmu.de}
\affiliation{Ludwig-Maximilians-Universit\"at M\"unchen, Fakult\"at f\"ur Physik, Schellingstra\ss e 4, 80799 M\"unchen, Germany}
\affiliation{Max-Planck-Institut f\"ur Quantenoptik, Hans-Kopfermann-Stra\ss e 1, 85748 Garching, Germany}

\begin{abstract}
Imaging the optical properties of individual nanosystems beyond fluorescence can provide a wealth of information. However, the minute signals for absorption and dispersion are challenging to observe, and only specialized techniques requiring sophisticated noise rejection are available. 
Here we use signal enhancement in a high-finesse scanning optical microcavity to demonstrate ultra-sensitive imaging. Harnessing multiple interactions of probe light with a sample within an optical resonator, we achieve a 1700-fold signal enhancement compared to diffraction-limited microscopy. We demonstrate quantitative imaging of the extinction cross section of gold nanoparticles with a sensitivity below \(1\,\text{nm}^2\), we show a method to improve spatial resolution potentially below the diffraction limit by using higher order cavity modes, and we present measurements of the birefringence and extinction contrast of gold nanorods.
The demonstrated simultaneous enhancement of absorptive and dispersive signals promises intriguing potential for optical studies of nanomaterials, molecules, and biological nanosystems.

\end{abstract}

\maketitle

\section*{Introduction}
Nanoscience strives for tools that enable the characterization and imaging of nanoscale objects.
Heterogeneity in particle shape, molecular morphology, and microenvironment tends to wash out intrinsic properties and calls for single-particle sensitive techniques.
The most common method, fluorescence microscopy, provides specific contrast and strong signals, but is limited to fluorophores which suffer from photobleaching, and does not provide information about the intrinsic optical properties of non-fluorescent samples.
Detecting single-particle signals beyond fluorescence is a true challenge for small objects, since the polarizability and the absorption cross section scale as \(a^3\) with system size \(a\). To achieve the required sensitivity, the imaging techniques demonstrated to date are carefully optimized to measure one single quantity by reducing measurement noise, implementing noise rejection techniques, and by signal averaging \cite{Yurt12}. This has enabled imaging of weak sample absorption e.g.\ by photothermal microscopy \cite{Boyer02,Cognet03} or direct absorption spectroscopy \cite{Arbouet04,Celebrano11}, as well as imaging of dispersive objects by interferometric scattering \cite{OrtegaArroyo12,Piliarik14}.
In a complementary approach, ultra-sensitive measurements can be realized by using signal enhancement within an optical cavity. Experiments such as dispersive sensing with microresonators \cite{Vollmer08,Zhu09,Vollmer12,Baaske14}, photothermal frequency shift spectroscopy \cite{Heylman14}, and cavity ringdown spectroscopy \cite{Berden00,Ye98} harness multiple round trips of light inside a cavity to enhance sensitivity. However, the experiments to date lack control over the relative position between the sample and the cavity mode, which, for most approaches, makes them incompatible with imaging and hinders quantitative analysis on a single-particle level.

In this work we report on a versatile approach that combines cavity enhancement with high-resolution imaging and provides high sensitivity for both sample absorption and dispersion simultaneously. It is based on an open-access optical microcavity \cite{Hunger10b,Toninelli10,Muller10,Greuter14,Dolan10} made of two highly reflective mirrors, which permits imaging a sample by raster-scanning it through a microscopic cavity mode.
Figure \ref{fig:Setup} (a) shows the basic setup used in our experiments. Multiple round trips of light between the mirrors accumulate loss and dispersive phase shifts caused by the sample. For a sample localized in the field maximum, the enhancement compared to a single pass amounts to \(4\,\mathcal{F}/\pi\), given by the number of reflections \(\mathcal{F}/\pi \cong 1/(1-R)\) and the increased intensity owing to the standing wave in the cavity. Here, \(\mathcal{F}\) is the cavity finesse and \(R\) the reflectivity of both mirrors. 
Notably, the same enhancement is available for absorption, scattering, fluorescence, and dispersive signals \cite{Motsch10,TanjiSuzuki11}.
At the same time, the overall signal scales inversely with the mode cross section \(\pi w_0^2\), such that a cavity that maximizes the figure of merit \(\mathcal{F} \lambda^2/w_0^2\) is desired.

In the following, we report three sets of measurements that demonstrate the potential of cavity enhancement for ultra-sensitive imaging: First, we image gold nanoparticles with superior sensitivity and quantitatively evaluate their extinction cross sections. Second, we introduce a method to improve spatial resolution by combining higher order cavity modes and show how this can be used to surpass the diffraction limit. Finally, we make use of the simultaneous enhancement of absorptive and dispersive signals and demonstrate measurements of the polarization dependent extinction and polarizability of gold nanorods, providing a way for the detailled characterization of the polarizability tensor of a nanoscale sample. 

\section*{Results}
\subsection*{Scanning cavity setup}
\begin{figure}
\includegraphics{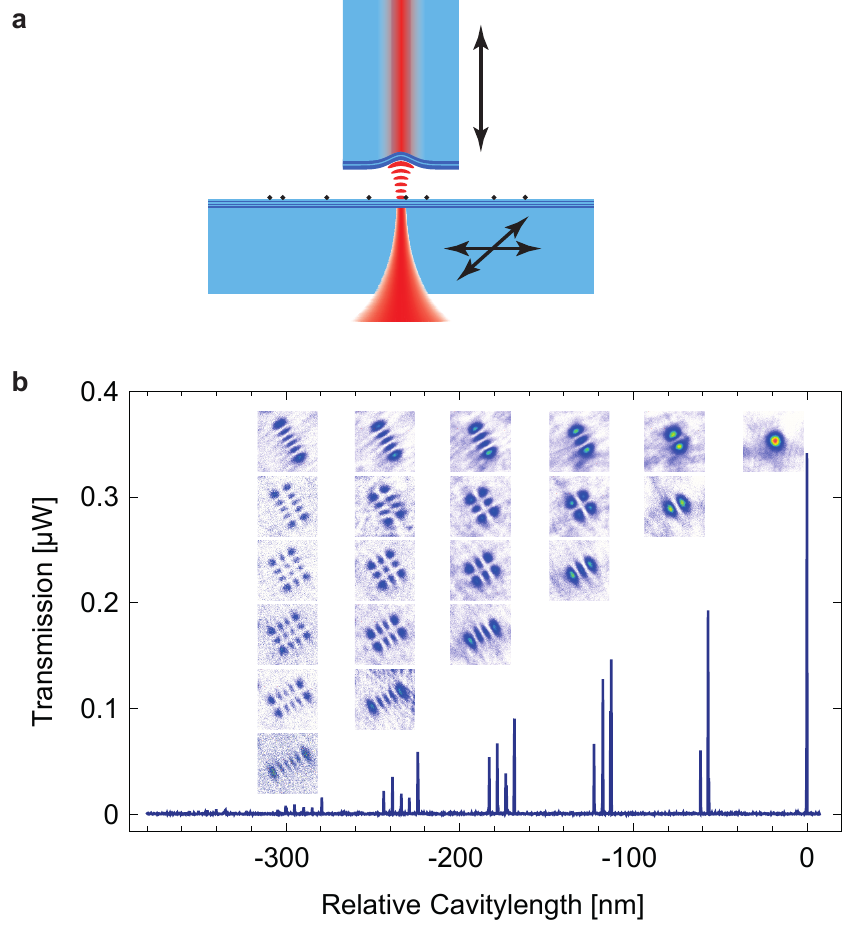}
\caption{
\label{fig:Setup}
\textbf{Schematic of a scanning cavity microscope.}
(a) A cavity built from a laser-machined and mirror-coated optical fiber and a planar mirror serving as a sample holder. Transverse scanning of the sample mirror is used for spatial imaging, axial scanning of the fiber for resonance tuning. (b) Cavity transmission signal when tuning the cavity length. The point spread functions of different transverse modes (insets) are measured by scanning the cavity across a single Au nanoparticle and evaluating the resonant transmission for each mode.
}
\end{figure}

In our approach, we use a high-finesse Fabry-Perot microcavity based on a mirror fabricated on the laser-machined endfacet of an optical fiber \cite{Hunger10b,Hunger12} (see Supplementary Figure 1). Combined with a scannable planar mirror serving as a sample holder, a cavity is formed that reaches a finesse of \(\mathcal{F}=57000\) and a mode waist of\ \(w_0 = 2.4\,\mu\text{m}\) for small mirror separations.

We employ a grating-stabilized diode laser at a fixed wavelength \(\lambda = 780\,\text{nm}\) to probe the cavity by tuning the mirror separation across a fraction of a free spectral range of the cavity with a high-precision closed-loop nanopositioner.
We detect the cavity transmission and observe several cavity modes becoming resonant at particular mirror separations. For each resonance, we evaluate the transmission, the linewidth, and its position. We use an electro-optic modulator to imprint sidebands on the laser as reference markers to correct for nonlinearities and mechanical noise (see Supplementary Figure 2).

Figure \ref{fig:Setup} (b) shows a typical detector raw transmission signal when the cavity length is scanned across a few hundred nanometers. The signal contains information about additional intracavity loss, which decreases the transmission and increases the linewidth, as well as about sample polarizability, which shifts the resonance position due to the associated effective refractive index change.
We raster-scan the sample mirror and record the transmission spectrum at each pixel, such that from a single measurement we can extract spatial images reflecting sample extinction and polarizability.

\subsection*{Extinction cross section of gold nanoparticles}
\begin{figure*}
\includegraphics{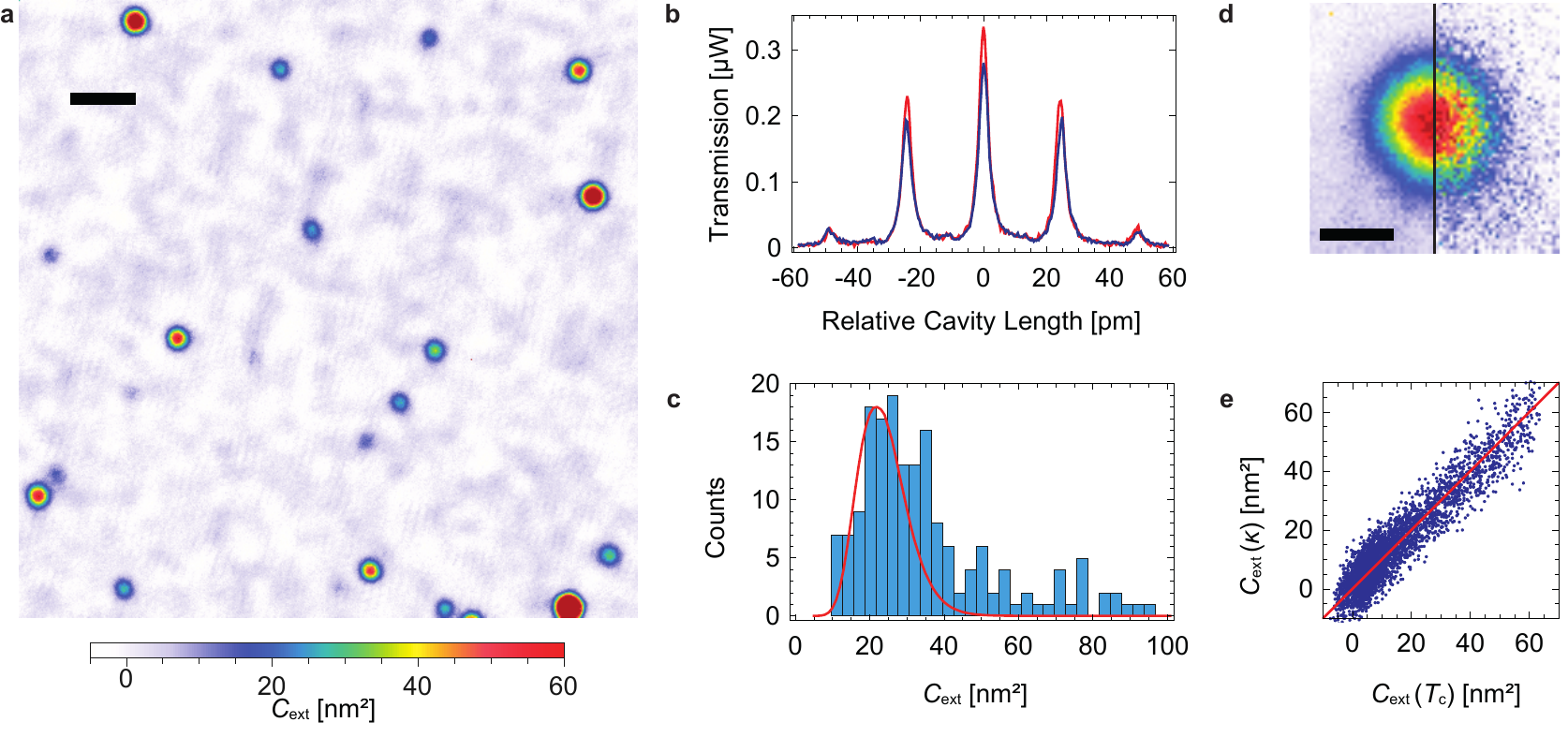}
\caption{
\label{fig:Gold}
\textbf{Extinction cross section of gold nanoparticles.}
(a) Spatially resolved map of the extinction cross section for a mirror carrying \(40\,\text{nm}\) gold nanospheres. Scale bars, \(10\,\mu m\).
(b)Transmission signal of the fundamental cavity mode with sidebands when centered on an individual nanosphere (blue) and on a clean mirror spot (red) .
(c) Histogram of the measured particle extinction cross section (blue) and calculated distribution (red solid line).
(d) Extinction measurement of a nanoparticle by transmission (left half) and linewidth (right half). Scale bars, \(1\,\mu m\)).
(e) Pixel-by-pixel comparison between the extinction cross section as measured by cavity transmission and linewidth. 
}
\end{figure*}

As a first step, we demonstrate quantitative imaging of particle extinction with high sensitivity. We study \(40\,\text{nm}\) gold nanoparticles as a reference system whose extinction cross section can be calculated. We choose a wavelength far away from the plasmon resonance, where the extinction cross section has dropped by nearly two orders of magnitude and amounts to \(\sim 2\%\) of the geometrical particle cross section.

Figure \ref{fig:Gold} (a) shows an example for a measurement where we evaluate the resonant cavity transmission of the fundamental mode. On resonance, the cavity transmission is given by
\(T_c = \epsilon_{m,n}\,4\,T_1 T_2/(T_1+T_2+L_1+L_2+2B)^2\).
Here \(T_i,L_i, i=\{1,2\}\) is the respective mirror transmission and loss, which can be inferred from measurements on a clean mirror with high precision (\(\sim 5\%\) uncertainty), and \(B\) is the additional loss introduced by the sample. The mode matching \(\epsilon_{m,n}\) between the fiber mode and the respective cavity mode with mode index \((m,n)\) can be tuned by angular alignment of the fiber with respect to the plane mirror to achieve controlled coupling to modes up to order \(m+n \sim 8\).
From the additional loss, we can quantitatively extract the extinction cross section of the sample,
\(C_\mathrm{ext} = B \pi w_0^2/4\). The mode waist of the cavity can be obtained from the point spread function (PSF) observed for a point-like particle, see figure \ref{fig:Gold} (d).

The spatial map shown in figure \ref{fig:Gold} (a) shows a large spread in the extinction cross section even for a monodisperse sample. To characterize this in detail, we determine the peak extinction cross section and the spot size of each loss feature and histogram the peak values for those features, whose size agrees with the size of the PSF of the cavity. This ensures that we mostly select only individual nanoparticles for the evaluation. We obtain a distribution peaking at \(C_\mathrm{ext}=25\,\text{nm}^2\) with a full width at half maximum of \(22\,\text{nm}^2\).
We compare the measurement to a calculation of the extinction cross section, where we take into account absorption and scattering \cite{Hulst57}, the effect of the mirror surface \cite{Wind87}, and surface-scattering induced damping \cite{Muskens08} (see Supplementary Note 1). For \(41\,\text{nm}\) gold particles on a fused silica surface at a wavelength of \(780\,\text{nm}\) we calculate \(C_\mathrm{ext} = 22\,\text{nm}^2\). 
Together with the specified size distribution, we obtain a calculated distribution which reproduces the data very well without any free parameter, see figure \ref{fig:Gold} (c). This underlines the potential of our technique for quantitative sample characterization.

In the same manner, we can also evaluate the resonance linewidth \(\kappa= c\, (T_1+T_2+L_1+L_2+2B)/(4\, d)\),
with \(c\) the speed of light and \(d\) the length of the cavity including penetration of the field into the dielectric mirror.
While the linewidth is immune to intensity fluctuations and thereby shows less drifts, the shot-to-shot variation is larger due to mechanical noise coupling to the cavity. When comparing the extinction extracted from linewidth and transmission data we observe the expected linear correlation with unity slope, see figure \ref{fig:Gold} (e).

We emphasize, that due to the cavity, the small extinction cross section of \(25\,\text{nm}^2\) leads to a large
(\(17\%\)) change in the raw transmission signal as shown
in figure \ref{fig:Gold} (b). This is in contrast to the expected single-pass signal of \(2\, C_\mathrm{ext}/(\pi w_\mathrm{DL}^2) = 1.0\times 10^{-4}\) for a diffraction-limited microscope, where \(w_\mathrm{DL} \approx \lambda/2\). In comparison, the cavity enhances the signal by a factor of 1700, in agreement with the expected factor of \(4\,\mathcal{F}/\pi (w_\mathrm{DL}^2/w_0^2)\).
In the measurements shown, we achieve a sensitivity for extinction cross sections of \(0.5\,\text{nm}^2\), limited by spatial variation of the background on a clean mirror (see Supplementary Figure 3). The
achieved sensitivity corresponds to the extinction cross section of \(2\,\text{nm}\) gold spheres at the plasmon resonance and is comparable to e.g. typical values for cross sections of single quantum emitters at room temperature \cite{Celebrano11}.

\subsection*{Resolution improvement with higher order transverse modes}
\begin{figure*}
\includegraphics{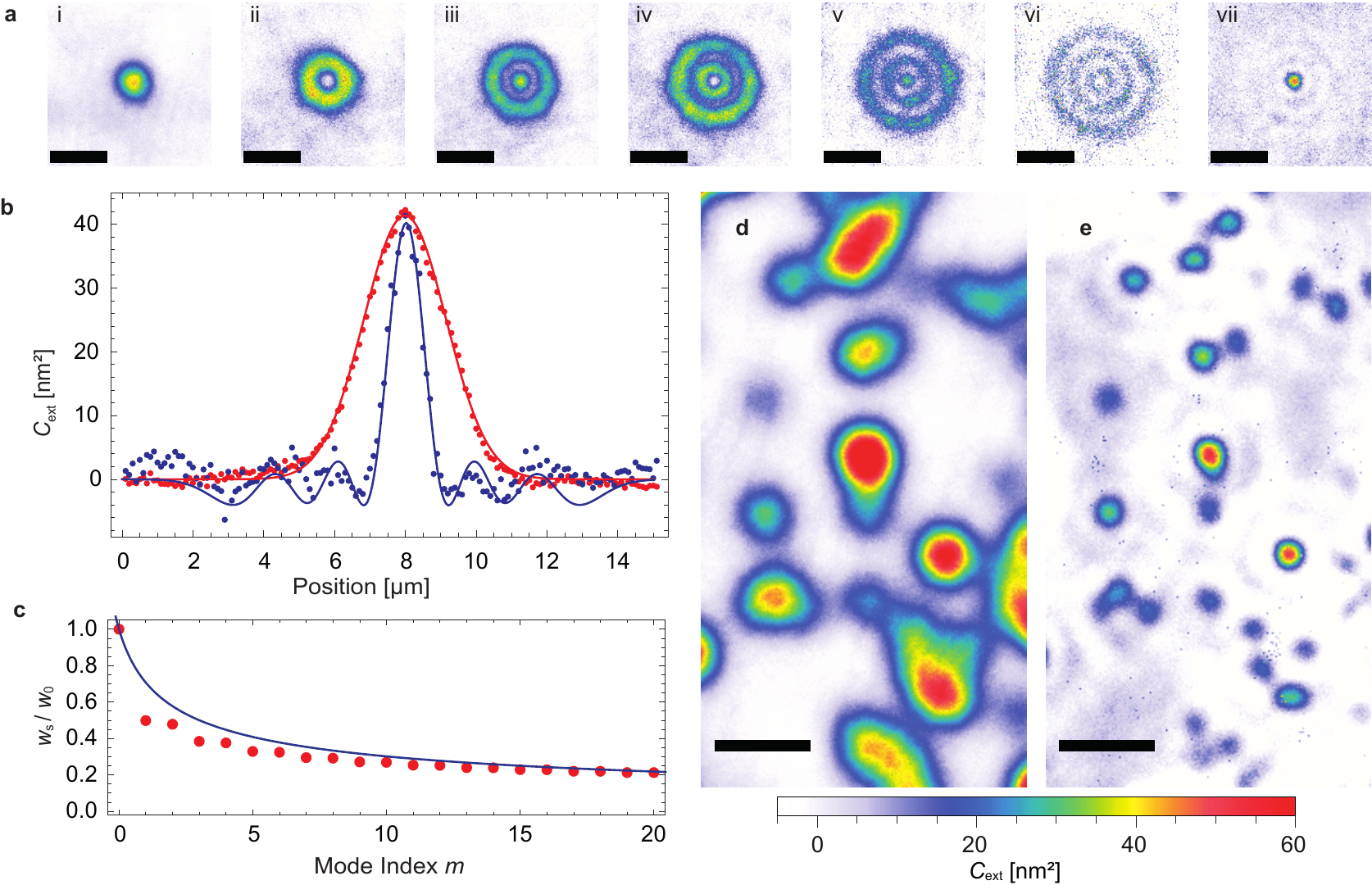}
\caption{
\label{fig:Superresolution}
\textbf{Resolution enhancement by higher transverse modes.}
(a) Extinction measurement of a single particle using the fundamental mode (i) and the first five higher transverse-mode families (ii) -- (vi), combined to yield an enhanced-resolution mode (vii). Scale bars, \(5\,\mu m\)
(b) Averaged section (5 rows) through the fundamental mode (red dots) and enhanced-resolution mode (blue dots) together with a fit (solid lines).
(c) Achievable spatial resolution improvement, showing the ratio of the reduced (\(w_s\)) and initial (\(w_0\)) mode waist as a function of the maximal mode order included.
(d) Extinction map of \(40\,\text{nm}\) Au NPs.
(e) Enhanced-resolution map of the same area using the higher mode orders up to \(m+n=3\). Scale bars, \(10\,\mu m\)
}
\end{figure*}

As a next step, we demonstrate a method to improve the spatial resolution by using higher order cavity modes \cite{Horak02, Dorn03,Gigan05}. Higher transverse modes carry larger transverse momentum and can thus be used to resolve smaller structures, similar to the concept used in structured illumination microscopy \cite{Gustafsson08}. Here, we follow the principle of constructing a squeezed state of the quantum harmonic oscillator,
where a suitable superposition of Hermite-Gaussian (HG) states yields squeezed states with reduced position uncertainty and correspondingly increased momentum uncertainty. We adopt this approach to optical cavity modes, and construct an effective mode with a spatial distribution that is smaller than the Gaussian fundamental mode. Since the HG modes are not simultaneously resonant due to the mode-dependent Gouy phase, only an incoherent superposition is possible, in contrast to the coherent superposition for squeezed states. Still, one can find a suitable expansion of the intensity distributions that closely approximates a squeezed state. Considering a 1D situation to illustrate the principle, we use
\begin{equation}
\label{eq:LinComb}
\Psi = \sum_{m=0}^\infty (-1)^m c_m(\rho) \left|\phi_m\right|^2,
\end{equation}
with coefficients
\(c_m(\rho) = m!/(2^{m}(m/2)!^2) \times \textrm{tanh}^{m}(\rho)/\textrm{cosh}(\rho)\) containing the "squeezing strength" \(\rho\), and the HG modes
\(\phi_{m} =  \mathcal{N}_{m}/w_0\times H_m(\sqrt{2}r/w_0) e^{-r^2/w_0^2}\)
with the Hermite polynomials \(H_m\) and the normalization
\(\mathcal{N}_{m} = \sqrt{\sqrt{2}/(2^mm!\sqrt{\pi})}\) \cite{Gerry04}.
We also include odd HG modes in a way that the linear combination adds the even and subtracts the odd modes. This has almost the same effect as the interference that is present for the coherent superposition.

We evaluate the localization of \(\Psi\) by inferring the position \(w_s\) where \(\Psi = 1/e^2\) for different numbers of HG modes contributing. At this stage, we remain in the paraxial approximation, and discuss deviations below. We find that the resolution improves according to \(\Delta x \approx \sqrt{1/(m_\mathrm{max}+1)}\), where \(m_\text{max}\) is the largest mode order included, see figure \ref{fig:Superresolution} (c). This is in accordance with the expected scaling that results from the increase of the number of transverse field nodes \(\propto m\) and the increase of the mode radii \(w_m \approx w_0\sqrt{m+1}\). In consequence, for an optical system where the numerical aperture (NA) is not fully used, resolution can be increased at least down to the diffraction limit. This is the case for optical microcavities, which can have a NA approaching unity for small mirror separation, but where the waist of the fundamental mode remains larger than the diffraction limit because the mirror radius of curvature is large compared to the mirror separation.

Figure \ref{fig:Superresolution} shows the experimental realization of this concept, which involves an extension of the above principle to 2D. In the measurements, we record the transmission and linewidth of all different modes within a single measurement by recording traces such as the one shown in Fig.\ \ref{fig:Setup} (b) for each pixel, where within a few microseconds, all modes are probed and recorded sequentially. As a first step, we evaluate the transmission of cavity modes up to the fifth order (\(m+n=5\)), comprising all 15 modes shown in figure \ref{fig:Setup} (b). We sum over all transverse modes belonging to one mode order, which leads to rotationally symmetric, concentric ring shapes, see figure \ref{fig:Superresolution} (a). Combining these modes according to equation \ref{eq:LinComb}, we arrive at the enhanced-resolution PSF shown in figure \ref{fig:Superresolution} (a, vii). We set \(\rho=2\), which at the same time optimizes the spatial resolution and minimizes oscillations of the outer part of the PSF. In figure \ref{fig:Superresolution} (b), an averaged section through \(\left|\phi_{00}\right|^2\) and \(\Psi\) as well as fits to the measured values using the described model are presented.
The enhanced-resolution PSF has an \(1/e^2\) radius of \(0.87\,\mu\text{m}\), a factor of 2.7 smaller than the fundamental mode, and a factor of 2.2 away from the diffraction limit of \(\lambda/2 = 390\,\text{nm}\).

Figures \ref{fig:Superresolution} (d) and (e) show the improved resolution when imaging \(40\,\text{nm}\) Au nanospheres in direct comparison. Features that are not resolvable with the fundamental mode become clearly separated, and the background is only weakly modulated due to the oscillations of the outer part of the PSF. At the same time, the high sensitivity, the enhancement factor, and the quantitative character of the measurement is maintained.

Due to its relevance for imaging in general, we briefly touch on the achievable limits of resolution when considering a rigorous calculation of the vector field without relying on the paraxial approximation \cite{Richards59,Novotny06}. We calculate the focal field of Hermite-Gaussian modes when projected from the far field. We find that only the \(\phi_{01 / 10}\) modes are useful for resolution enhancement, since higher modes are not suited for input aperture overfilling. A calculation of \(\Psi = \left|\phi_{00}\right|^2-0.6(\left|\phi_{01}\right|^2+\left|\phi_{10}\right|^2)\) yields an effective PSF with a first zero crossing at \(\Delta x \approx 0.39 \lambda/\mathrm{NA}\), such that the resolution can be improved by a factor 1.5 below the diffraction limit \(\Delta x \approx 0.61 \lambda/\mathrm{NA}\). The scheme could be easily implemented in standard confocal microscopes by using two illumination paths, providing both standard illumination for a (near) Gaussian spot, and illumination with an azimuthally polarized doughnut mode. Acquisition of two images and subsequent subtraction provides superresolution with little overhead also for non-fluorescent objects.

\subsection*{Extinction contrast and birefringence of Au nanorods}
\begin{figure*}
\includegraphics{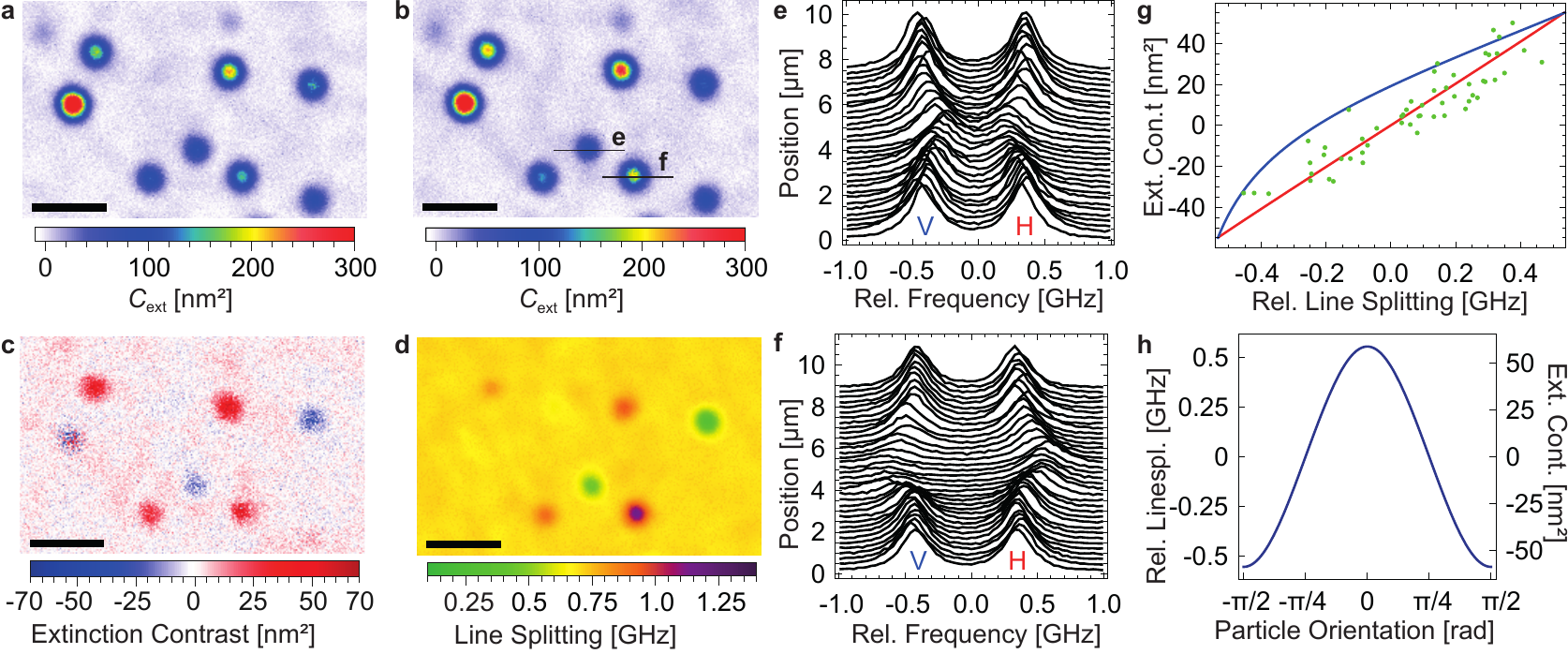}
\caption{
\label{fig:Polarization}
\textbf{Extinction contrast and birefringence imaging.}
Extinction map of Au-nanorods for \(H\) - polarized  (a) and \(V\) - polarized (b) light.
(c) Extinction contrast \(C_\mathrm{ext}^V-C_\mathrm{ext}^H\).
(d) Absolute splitting of the two orthogonally polarized fundamental cavity modes. Scale bars, \(10\,\mu m\).
(e), (f) Transmission signal of the cavity while scanning over the nanorods denoted in (b).
(g) Correlation between relative line splitting and extinction contrast. Measured values (green dots), calculated correlation assuming a fixed (red solid line) and variable (blue solid line) cavity mode orientation are shown.
(h) Relative line splitting and extinction contrast in dependence of the orientation of the nanorod.}
\end{figure*}
More detailed information about the optical properties can be obtained when studying both absorption and dispersion of a sample. Furthermore, many samples of interest lack spherical symmetry, and their optical response depends on the relative orientation between the laser polarization and the eigenaxes of the samples' polarizability tensor. The resulting angle dependent extinction and dispersion give rise to observable extinction contrast and birefringence. However, for nanoscale samples, the signals are minute and their simultaneous imaging has been out of reach so far. Here we show that by monitoring resonance frequency shifts of the cavity, we can image dispersive properties of a sample in parallel with sample extinction in a polarization-sensitive way.

As an example, we study gold nanorods of size \(34 \times 25\times 25~\mathrm{nm}^3\), which are expected to show extinction contrast and birefringence due to their anisotropic shape. Given the cylindrical symmetry, the complex polarizability tensor simplifies to the components parallel and perpendicular to the long axis, \(\{\alpha_\parallel,\alpha_\bot\}\). We demonstrate simultaneous imaging of the extinction contrast and birefringence by measuring the linewidths and frequency splitting of two orthogonally polarized cavity modes. This provides information about the orientation and shape and thus the internal structure of the sample.

In our experiment, we have to take into account the intrinsic mode splitting present in our cavity. Due to ellipticity of the laser-machined mirror surface profiles, the modes of the cavity are split into a linear polarization doublet (denoted by \(H \rightarrow\; \text{fast}\), \(V \rightarrow\; \text{slow}\)), whose axis and splitting is determined by the mirror shape \cite{Uphoff15}.
By setting the input polarization, we can excite both modes as shown in figure \ref{fig:Polarization} (e, f), and evaluate their response to study polarization effects.

Figure \ref{fig:Polarization} (a, b) shows the extinction cross section for the two polarization modes as inferred from the cavity linewidth. We evaluate the difference of the extinction cross section \(C_\mathrm{ext}^V-C_\mathrm{ext}^H\) and find significant values for most of the particles, see figure \ref{fig:Polarization} (c).
Figure \ref{fig:Polarization} (h) displays the expected values for \(C_\mathrm{ext}^V-C_\mathrm{ext}^H\) as a function of the particle orientation, showing good agreement with the observed range (see Supplementary Note 1).

From the same measurement, we infer the birefringence of the sample by monitoring the separation of the polarization modes of the cavity. 
The differential frequency shift \(\Delta \omega\) of the cavity resonance doublet due to a birefringent particle is found to be \cite{Arnold03,Noto07}
\begin{equation}
\frac{\Delta \omega}{\kappa} = -\frac{\mathrm{Re}(\alpha_V-\alpha_H)}{\epsilon_0}\frac{4\,\mathcal{F}}{\pi w_0^2\lambda},
\end{equation}
where we have normalized to the bare cavity linewidth \(\kappa\) and used the projection of the polarizability eigenaxes onto the cavity eigenaxes. For a nanorod with the long axis in the plane of the mirror, this is given by \(\alpha_H^2 = \alpha_\bot^2\cos^2\theta + \alpha_\parallel^2\sin^2\theta\), \(\alpha_V^2 = \alpha_\bot^2\sin^2\theta + \alpha_\parallel^2\cos^2\theta\), with the angle \(\theta\) between the orientation of the long particle axis and the \(V\) cavity mode.

Figure \ref{fig:Polarization} (d) shows a spatial map of the measured birefringence signal with the constant offset of \(740\,\text{MHz}\) originating from the intrinsic cavity birefringence. Figure \ref{fig:Polarization} (h) displays a calculation of the relative cavity mode splitting in presence of a gold nanorod as a function of particle orientation, where we assume a polarizability volume \(\mathrm{Re}(\alpha)/(4\pi\epsilon_0) = 3.3\times 10^{-17}\,\text{cm}^3\) (\(2.1\times 10^{-17}\,\text{cm}^3\)) for the long (short) axis (see Supplementary Note 1).

Figures \ref{fig:Polarization} (e, f) show two example signatures of nanorods that are mostly orthogonal (parallel) to the slow cavity eigenaxis, and thereby reduce (increase) the intrinsic birefringence. At the same time, the birefringence is correlated with the extinction difference, where the mode parallel to the long particle axis is affected more strongly.

In Fig. \ref{fig:Polarization} (g), we show the evaluated extinction difference and the relative dispersive frequency shift of a large number of nanorods. We observe good agreement between the expected and measured range of values. The correlation of the two quantities shows a linear relation and agrees with the prediction, which is again calculated without free parameters. With \(C_\mathrm{ext}^{H,\,V} \propto \Im(\alpha_{H,\,V})\) and \(\Delta \omega\propto \mathrm{Re}(\alpha_{H,\,V})\), the correlation provides a detailed characterization of the polarizability tensor of the sample.
Notably, the data yields good agreement with the expected signal when assuming a fixed orientation of the cavity eigenaxis. This is in contrast to the expectation that the cavity eigenmodes are rotated by the presence of sample birefringence when \(\theta\neq 0\) \cite{Brandi97,Moriwaki97}. The absence of rotation is furthermore confirmed by evaluating the cavity transmission after a polarizing beam splitter (see Supplementary Figure 4). This suggests that the geometry-induced mode splitting fixes the eigenmode axes.

In the measurements on line splitting, we achieve a noise floor of \(9\,\text{MHz}\) rms,
yielding a sensitivity for a polarizability volume difference of \(\mathrm{Re}(\alpha_V-\alpha_H)/(4\pi\epsilon_0)=2\times 10^{-19}\,\text{cm}^3\). Since we probe the nanorods far away from the plasmon resonance, their polarizability is comparable to the value of dielectric objects of same size. The demonstrated sensitivity should thus allow spatial imaging of e.g.\ individual macromolecules with a size down to a few tens of nanometers.

\subsection*{Discussion}
We have demonstrated a versatile technique for sensitive optical imaging based on an open-access, scannable microresonator. The combination of high spatial resolution, simultaneous high sensitivity for absorption and birefringence, and the quantitative nature of the signals promises great potential for label-free biosensing, characterization of nanomaterials, particle sizing, and spectroscopy of quantum emitters on a single particle level.

Additionally, our method could provide new insight into the microscopic properties of low loss mirrors, as used e.g.\ for gravitational wave detectors, cavity QED experiments, and laser gyroscopes.
Finally, fluorescence nanoscopy methods such as STED microscopy \cite{Hell07} could be implemented, where the near-ideal shape of the higher order modes, the intrinsic power enhancement, and Purcell enhancement of spontaneous emission could add significant benefit.

Our method is still open for substantial improvements of the sensitivity. The current limitation due to spatial background variation could be overcome by differential measurements, e.g.\ before and after application of the sample. Furthermore, the signal enhancement can be further increased by improved cavities, where we expect mode waists $w_0 < \lambda$ and a finesse \(\mathcal{F} > 200 000\) to be achievable.
Additionally, our method could be combined with noise reduction techniques, such that the relative noise level of \(2\times 10^{-2}\) reached in most of the measurements shown here, could be reduced potentially down to \(10^{-6}\) \cite{Ye98,Celebrano11}.

\section*{Methods}
\setlength{\leftmargini}{0pt}
\begin{description}
\item[Experimental Setup]
The microcavity is based on a micromirror on the endfacet of an optical fiber, which has a laser-machined concave depression with an effective radius of curvature of \(60\,\mu\text{m}\). It is coated with a dielectric mirror with \(R=99.9976\%\).
A planar \(1/2"\) mirror with \( R=99.9914\%\) serves as a sample holder. The cavity is typically operated at a length of \(d \approx 30 \,\lambda /2\) to avoid transverse-mode coupling \cite{Klaassen05,Benedikter14}, where it has a linewidth of \(245\,\text{MHz}\) and corresponding quality factor \(Q = 1.57\times10^6\).
The cavity is probed with an external cavity diode laser at a wavelength of \(780\,\text{nm}\) with a linewidth below \(1\,\text{MHz}\) (Toptica DL pro), coupled to the cavity fiber. The laser is phase-modulated with an EOM (Newfocus 4221) to generate sidebands used as frequency markers for linewidth and line splitting measurements. We excite the cavity such that the power circulating inside the cavity remains smaller than \(10\,\text{mW}\). The transmitted light is split up by a polarizing beam splitter aligned along the polarization axes of the resonator. The light is detected with two photodiodes (Thorlabs APD120) and recored with an oscilloscope (LeCroy HRO66Zi). The laser intensity can be adjusted for each cavity mode order with an AOM. The length of the cavity is scanned with a closed-loop piezo-linear actuator (Physikinstrumente LISA P-753, E-712).
Its absolute length is determined from white light transmission spectra, where we use a superluminescent diode (EXALOS EXS7505-8411) and analyze the transmitted light with a spectrometer (Ocean Optics HR4000). The length is stabilized by comparing the resonance positions of the \(780\,\text{nm}\) probe laser and a homebuilt external cavity diode laser at a wavelength of \(969\,\text{nm}\), where the cavity has a finesse below 1.
Both lasers are intensity but not frequency stabilized.

The \(1/2"\) plane mirror is scanned transversally with a xy piezo scanner (Physikinstrumente P-734, E-712).
A sketch of the optics can be seen in supplementary figure 1.

\item[Data analysis]
The recorded cavity resonances are fitted online with a sum of Lorentzians. The linewidth and the line splitting is determined by using sidebands modulated on the laser as frequency markers.
The measurement time amounts to \(\sim 100\,\text{ms}\) per pixel for the data shown, including the time needed for positioning, the length scan, data transfer, and data evaluation. In principle, the time response of the cavity is the fundamental limiting factor, with a typical time constant of \(10\,\text{ns}\). We expect that with a cavity locked on resonance while scanning transversally, this limit can be approached.

The transmission data are normalized such that the most probable value for an empty mirror is 1. The normalized data is then rescaled with the calculated transmission of an empty cavity.
For evaluating the amplitudes of the extinction cross section or line shift of spatial images, we fit a Gaussian to each circular object. For further analysis, we select only particles who's image size corresponds to the PSF, in order to exclude larger particles or clusters of particles.

\item[Nanoparticles and sample preparation] We spincast \(100\,\mu\text{l}\) of a 1:33 dilution of colloidal gold nanospheres with a diameter of 40 nm (British Biocell International, plasmon resonance at \(530\,\text{nm}\)) or colloidal nanorods with a size of \(25 \times 25 \times 34~\text{nm}^3\) (Strem Chemicals, plasmon resonance at \(550\,\text{nm}\)) onto a clean cavity mirror. Supplementary Figure 5 shows a scanning electron microscope image of a sample of gold nanospheres.
\end{description}

\section*{Acknowledgements}
We thank H. Kaupp and C. Deutsch for assistance in fabricating the cavity fiber, T. H{\"u}mmer for contributions during the early stage of the experiment, and Q. Unterreithmeier for assistance with SEM measurements.

This work was partially funded by the excellence cluster Nanosystems Initiative Munich (NIM).
T. W. H. acknowledges funding from the Max-Planck Foundation.

\section*{Author Contributions}
M. M. built the experiment, conducted the measurements, analyzed the data, and wrote the article.
D. H. devised and planned the experiment, analyzed the data, and wrote the article.
J. R. and T.W.H. discussed the experiment and the results, and contributed to writing the paper.

\section*{Additional Information}
The authors declare no competing financial interests.

\newpage

\onecolumngrid
\appendix

\renewcommand{\figurename}{Supplementary Figure}
\renewcommand{\refname}{Supplementary References}
\setcounter{equation}{0}

\section*{Supplementary Figure 1: Micro-Machined Fiber}
\begin{center}
\includegraphics{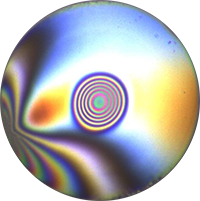}
\end{center}
Endfacet of the laser-machined fiber used in the experiment. Picture taken with a home-buit white-light interferometric microscope. The fiber endfacet is mirror coated with a commercial IBS coating (R=99.9976\%, T=L=12\,ppm) from ATF, Boulder.

\newpage

\section*{Supplementary Figure 2: Scanning Cavity Setup}

\begin{center}
\includegraphics{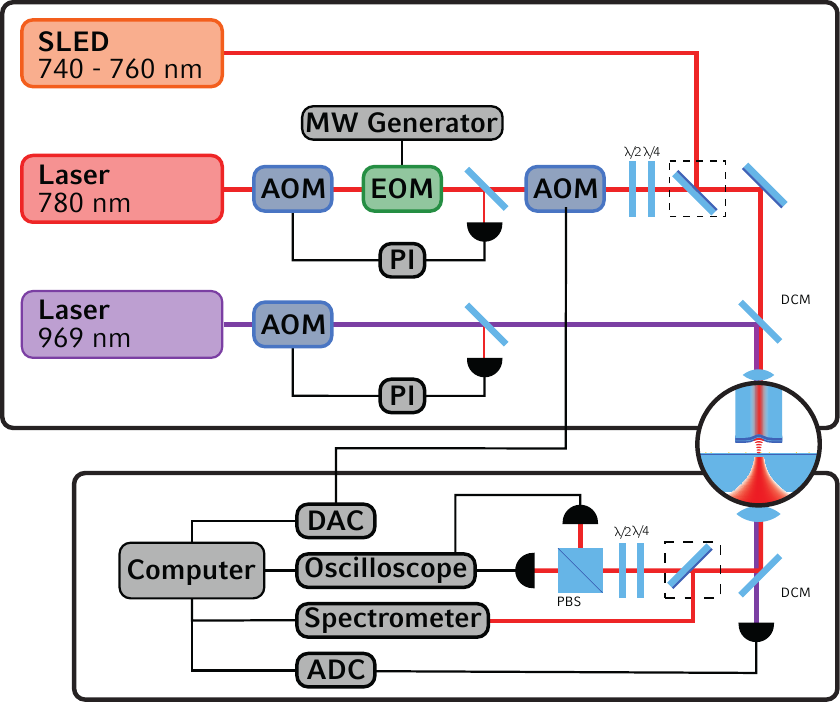}
\end{center}

\paragraph* {Probe laser}
The cavity is probed by a grating stabilized external cavity diode laser (TOPTICA DLpro) at a wavelength of 780 nm. The power of the laser ist stabilized to a relative noise level of \(10^{-4}\) by a feedback loop consisting of an AOM, a PI controller and a photodiode.

The laser is phase-modulated with a resonant EOM (Newfocus 4421) at a frequency of 1.748 GHz to generate sidebands which are used as frequency markers for cavity linewidth measurements as well as for measurements of the linesplitting of the two orthogonally polarized cavity modes.

As different transverse modes of the cavity have significantly different mode matching to the input light, the power of the probe light can be adjusted by a second AOM. It is possible to switch the intensity in a way that all recorded resonances have nearly the same intensity on the detectors in order to fully use the dynamic range of the detector and the oscilloscope.

The transmitted light is collected by an achromatic lens. The two orthogonally polarized modes of the cavity are separated by a polarizing beamsplitter. The light is detected by two APD's (Thorlabs APD120A). The signal is then recored by an oscilloscope (LeCroy HRO66Zi).

\paragraph*{Cavity length measurement}
To measure and adjust the length of the cavity precisely, the cavity can be probed with a superluminescent LED (EXALOS EXS7505-8411). By analyzing the transmitted light on a spectrometer (Ocean Optics HR4000), we can deduce the absolute length of the cavity including the field penetration into the dielectric mirrors from the spectral spacing of subsequent resonances.

\paragraph{Stabilizing the cavity length}
While scanning the surface of the plane mirror, the cavity length can change because of thermal and mechanical drifts, or a wedged plane mirror. If the cavity length changes, the relative position of the resonances compared to the beginning of the cavity length scan changes, and thus the resonances of interest can shift out of the data recording window.

To compensate this, we use a second laser (homebuilt external cavity diode laser) at a wavelength of 969 nm. At this wavelength the finesse of the resonator has dropped below 1. Thus, the resonator is not sensitive to weak absorbers. By comparing the actual position of a cavity resonance with a target value, we can readjust the start position of the length scan and thereby stabilize the average length of the cavity to within a few nanometers over days.

\newpage

\section*{Supplementary Figure 3: Plane Mirror}
\begin{center}
\includegraphics{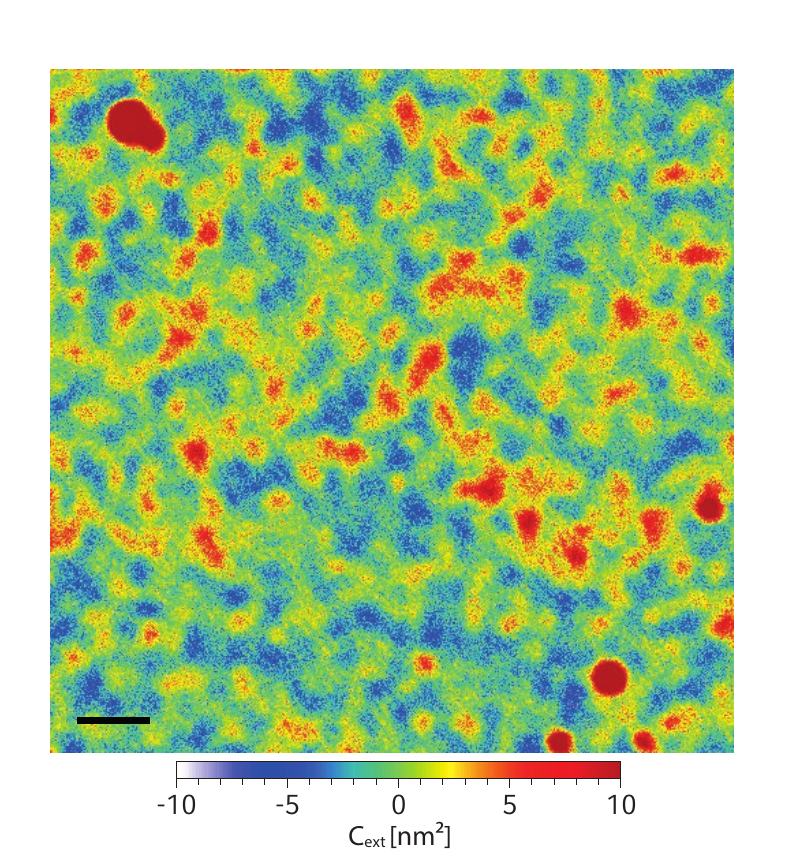}
\end{center}

The plane mirrors used as sample holders in our experiments are superpolished fused silica substrates, coated with a high reflective dielectric coating (R=99.9914\%, T=60\,ppm, L=26\,ppm) with SiO$_2$ as terminating layer, while the back side is anti-reflection coated. The coatings were manufactured by LAYERTEC, Mellingen.

The figure shows a map of extinction cross sections measured for a clean plane mirror. The typical variation of the mirror loss amounts to \(0.2\,\text{ppm}\;\text{rms}\), corresponding to an extinction cross section of \(0.5\,\text{nm}^2\) for the cavity parameters used in the experiments.

\newpage

\section*{Supplementary Figure 4: Birefringence of a Cavity}
\begin{center}
\includegraphics{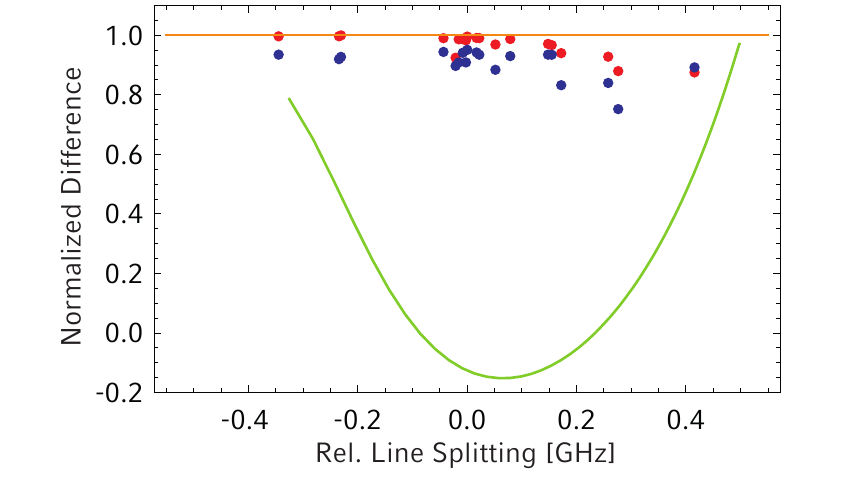}
\end{center}

We observe that each mode of the cavity is split up into two orthogonally polarized modes. Uphoff et al. \cite{Uphoff15} show that the small ellipticity of the micromachined mirror on the fiber endfacet leads to a differential phaseshift between light polarized along the two half axes of the resonator. Thus each mode of the cavity is split up into a doublet.
We exploit this for measuring anisotropy of extinction and cavity lineshift of single gold nanorods. For our theoretical considerations, we calculate the extinction cross section and the cavity line shift using a projection of the polarizability tensor of the particle onto the corresponding polarization axis.

We do not observe a rotation of the orientation of the cavity eigenaxes in the presence of particle birefringence, as predicted e.g. by Moriwaki et. al. \cite{Moriwaki97} or Brandi et. al. \cite{Brandi97}. If rotation would be induced, a second resonance peak would appear at the output ports of the polarizing beamsplitter, which is aligned with respect to the unperturbed cavity modes.

In order to prove this effect, we first fit the sum of both ports to get the position and width of each peak. In a second step, we fit each signal separately with a double Lorentzian, using the positions and widths from the first step in order to get the amplitudes of the two peaks, if they would appear.

The figure shows the measured difference of the two amplitudes of peaks for the two output ports of the beamsplitter. The values are normalized to the sum of both amplitudes. The data agree quite well with a model assuming no change of the polarization inside the cavity, while they obviously deviate from a model that assumes the nanoparticle changing the polarization orientation inside the cavity.

\newpage

\section*{Supplementary Figure 5: SEM Image of Gold Nano-Spheres}
\begin{center}
\includegraphics[width=\textwidth]{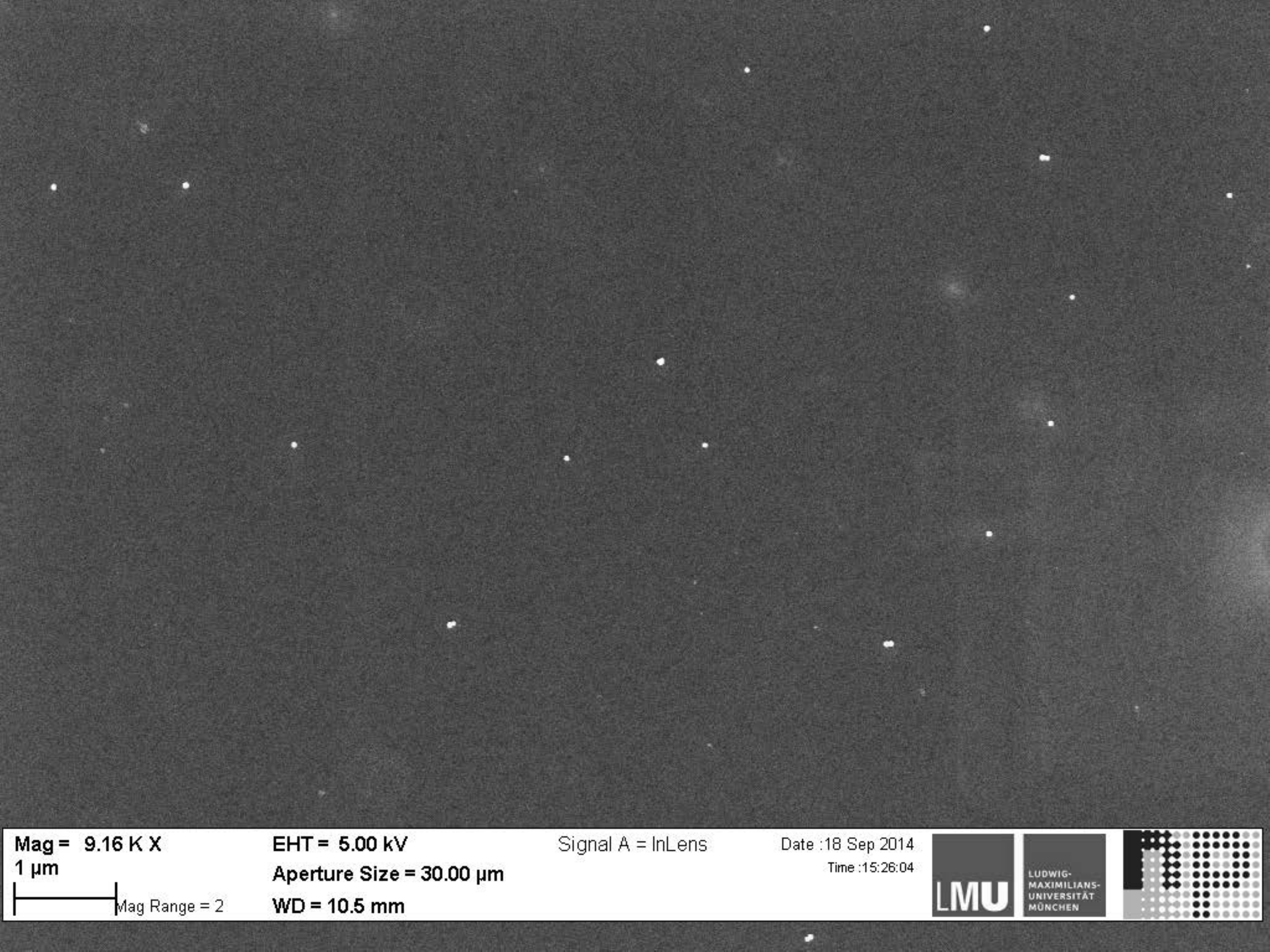}
\end{center}
Scanning electron microscope image of mirror with gold nanospheres used in the experiments.

\newpage

\setcounter{figure}{5}

\section*{Supplementary Note 1: Optical Properties of a Metal Nanoparticle}
For the interpretation of the experiments shown in this work, a detailed understanding of the optical properties of the metal nanoparticles is crucial. In the following, different effects that influence the properties are discussed.

The dielectric constants of the particle, its surrounding and the surfaces in the vicinity of the particle are denoted as follows: \(\epsilon_1\) denotes the dielectric constant of the surrounding environment, \(\epsilon_2\) the dielectric constant of a surface close to the particle and \(\epsilon_3\) gives the dielectric constant of the particle it self.
\(\epsilon_0\) is the vacuum dielectric constant. 
\subsection*{Polarizability of a nanoparticle}
Polarizability is the key quantity to describe the optical properties of a particle. In the following, basic formulas are presented:

\paragraph*{Spherical particles}
For spherically symmetric and isotropic particles, the polarizability is given by the Clausius-Mossotti law:
\begin{equation}
\alpha=3\epsilon_0 V_p \frac{\epsilon_3-\epsilon_1}{\epsilon_3+2\epsilon_1}
\end{equation}
where \(V_p\) is the volume of the particle.

\paragraph*{Ellipsoids}
For ellipsoidal nanoparticles, the polarizability is described by a tensor. The main values of the tensor can be calculated as follows \cite{Hulst57}:
\begin{equation}
\alpha_j=\frac{V_p \epsilon_0}{L_j+\frac{1}{\epsilon_3-1}},
\end{equation}
where \(j\) denotes the axis, \(V_p\) the volume of the particle and \(L_j\) is  a dimensionless parameter depending on the ratios of the semiaxes \(a_1\), \(a_2\) and \(a_3\) of the particle.
For arbitrary ratios of those axes, \(L_1\) is given by
\begin{equation}
L_1=\int_0^{\infty} \mathrm{d}s\frac{a_1\, a_2\, a_3}{2(s+a_1^2)^{3/2}\, (s+a_2^2)^{1/2}\,(s+a_3^2)^{1/2}},
\end{equation}
\(L_2\) and \(L_3\) are obtained by cyclical permutation. Furthermore, the relation
\begin{equation}
L_1+L_2+L_3=1
\end{equation}
has to be fulfilled.

For prolate ellipsoids like nanorods, where \(a_1>a_2,\,a_3\) and \(a_2=a_3\), the problem simplifies to
\begin{equation}
e^2=1-\frac{a_2^2}{a_1^2}, \quad L_1=\frac{1-e^2}{e^2}(-1+\frac{1}{2e}\ln\frac{1+e}{1-e}).
\end{equation}

For nanorods that are not aligned along the polarization of the probe light, we rotate the polarizability tensor and project it onto the polarization axes. We do the projection for the imaginary and real part separately.

\subsection*{Correction of the dielectric constant because of the small size}
It turns out that the dielectric constant of gold \(\epsilon_3\) has to be modified with respect to the bulk value (\(\epsilon_{bulk}\)) \cite{Johnson72} in order to describe very small particles accurately \cite{Muskens08}:
\begin{equation}
\label{eq:correction_epsilon}
\epsilon_3(\lambda)=\epsilon_{bulk}(\lambda)+\frac{\omega_p^2}{\omega(\omega+ i \gamma_{bulk})}-\frac{\omega_p^2}{\omega(\omega + i \gamma)},
\end{equation}
with the metal plasma frequency \(\omega_p\) and the electron scattering rates in the bulk (\(\gamma_{bulk}\)) and confinded metal (\(\gamma\)), where
\begin{equation}
\gamma=\gamma_{bulk}+2  g v_F / D
\end{equation}
with \(v_F\) the Fermi velocity, \(D\) the particle diameter and \(g\) a proportionality factor in the order of 1.

At a wavelength of 780 nm and $g=1.4$ we find an extinction cross section of  \(22.3\, \mathrm{nm}^2\).

\subsection*{Correction of the polarizability for particles on a dielectric surface}
As the nanoparticles in our experiments lie on the surface of a dielectric mirror, also the effect of a surface close to the particle has to be considered. For metal nanospheres far away from plasmon resonance and for an incident light field with a polarization parallel to the surface, like in the case discussed in this paper, Wind et al. \cite{Wind87} give the following solution:
\begin{equation}
\alpha=\epsilon_0 \frac{\epsilon_1(\epsilon_3-\epsilon_1) V_p}{\epsilon_1+L_s(\epsilon_3-\epsilon_1)}
\end{equation}
with the depolarization factor
\begin{equation}
L_s=\frac{1}{3}\left(1-\frac{1}{8} \frac{\epsilon_2-\epsilon_1}{\epsilon_2+\epsilon_1} \right).
\end{equation}

\subsection*{Scattering and Absorption}
When light interacts with an object it either gets absorbed or scattered. For nanoparticles smaller than a wavelength of light, this is described by Rayleigh scattering.
The corresponding cross sections are given by:
\begin{equation}
C_{abs}=\frac{2\pi \sqrt{\epsilon_1}}{\lambda \epsilon_0}\Im{(\alpha)}
\end{equation}
and
\begin{equation}
C_{sca}=\left(\frac{2\pi}{\lambda}\right)^4\frac{\left|\alpha\right|^2}{6\pi\epsilon_0^2}.
\end{equation}

As both effects attenuate the probe beam, the measured quantity in all our experiments is the extinction of a nanoparticle:
\begin{equation}
C_{ext}=C_{abs}+C_{sca}.
\end{equation}

\begin{figure}[htb]
\begin{center}
\includegraphics{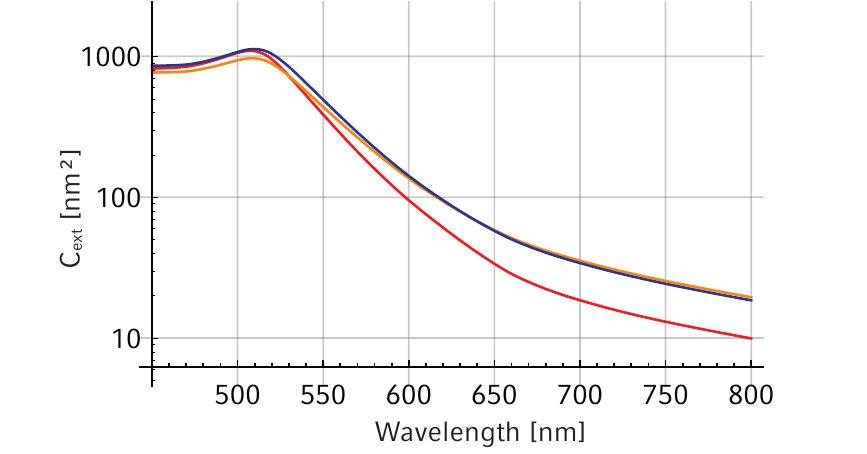}
\end{center}
\caption{Extinction spectrum of a gold nanosphere with a diameter of 40 nm. Particle in free space (air, vacuum) using the bulk value for \(\epsilon_3\) (red). Particle in free space, using the corrected value of \(\epsilon_3\) for small particles with \(g=1.4\) (orange). Particle on a fused silica surface, using the corrected \(\epsilon_3\) (blue).}
\label{figsup:extinction_wavelength}
\end{figure}

Supplementary figure \ref{figsup:extinction_wavelength} shows the extinction spectrum of a spherical gold nanoparticle with a diameter of 40 nm taking into account the effects discussed in this section.

\subsection*{Size- and g-factor distribution for gold nanospheres}
The size of the nanoparticles varies as well as the  \(g\)-factor, a heuristic value characterizing damping effects for small nanoparticles. In order to compare our measured distribution of extinction cross sections, we take a Gaussian diameter distribution measured by the manufacturer of the particles with central diameter \(41.1\,\text{nm}\) and a standard deviation of \(3.3\,\text{nm}\). In addition, we adopt a gaussian distribution of the \(g\)-factor found by Muskens et. al. \cite{Muskens08} around 1.4 with a standard deviation of 0.25.

\subsection*{Calculations for nanorods}
For all calculations concerning the polarizability of nanorods, we use the calculus described above for ellipsoids. We correct \(\epsilon_3\) for small-size effects of the particle. In equation \ref{eq:correction_epsilon}, we use the size of the long (short) axis as particle diameter for the calculation of the polarizability along the respective axis.
As we are interested in differences, we do not consider surface effects, which we assume to be equal for both axes.

\begin{figure}[htb]
\begin{center}
\includegraphics{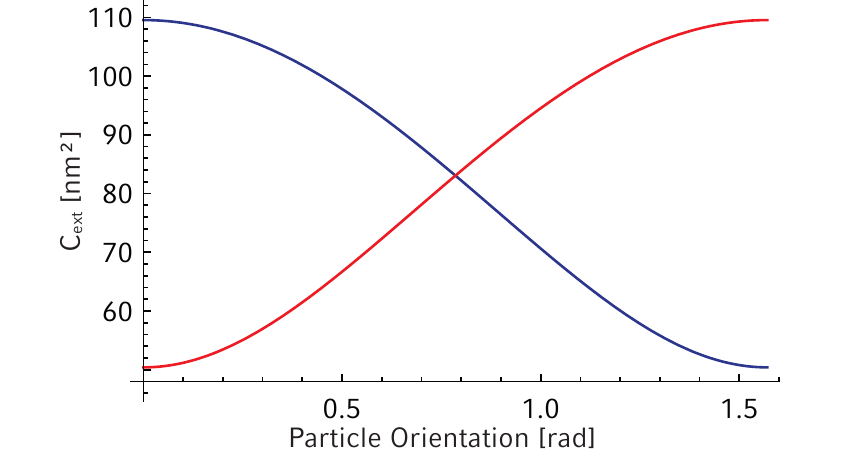}
\end{center}
\caption{Extinction cross section of a gold ellipsoid. The red and blue curve indicate the extinction for two orthogonally polarized light fields. At an angle of 0, the particle is aligned along its main axes.}
\label{figsup:extinction_ellipsoid}
\end{figure}

Supplementary figure \ref{figsup:extinction_ellipsoid} shows the calculated extinction cross section of a nanorod for two orthogonally polarized light fields depending on its orientation with respect to the polarization of the light field.

\end{document}